\documentclass[aps,a4paper,superscriptaddress,preprintnumbers,showpacs,tightenlines]{revtex4}
\usepackage{graphicx}
\usepackage{bm}

\begin{document}

%                      PRD Draft Version 2.2

\title{Study of three-body charmless $B$ decays}

\collaboration{The Belle collaboration}

\author{
  A.~Garmash$^{2,9}$,         % BINP+KEK
  K.~Abe$^{9}$,               % KEK
  K.~Abe$^{41}$,              % TohokuGakuin
  N.~Abe$^{44}$,              % TIT
  T.~Abe$^{42}$,              % Tohoku
  I.~Adachi$^{9}$,            % KEK
  H.~Aihara$^{43}$,           % Tokyo
  Y.~Asano$^{48}$,            % Tsukuba
  T.~Aso$^{47}$,              % Toyama
  V.~Aulchenko$^{2}$
% ...
\vspace*{0.5cm}\\
\small
% ...
$^{2}${Budker Institute of Nuclear Physics, Novosibirsk}\\
% ...
$^{9}${High Energy Accelerator Research Organization (KEK), Tsukuba}\\
% ...
$^{41}${Tohoku Gakuin University, Tagajo}\\
$^{42}${Tohoku University, Sendai}\\
$^{43}${University of Tokyo, Tokyo}\\
$^{44}${Tokyo Institute of Technology, Tokyo}\\
% ...
$^{47}${Toyama National College of Maritime Technology, Toyama}\\
$^{48}${University of Tsukuba, Tsukuba}\\
% ...
}

\begin{abstract}

  We report on a study of three-body charmless decays $B^+\to K^+h^+h^-$ 
based on a $29.1$~fb$^{-1}$ data sample collected with the Belle detector.
With no assumptions on the intermediate mechanisms, the following 
three-body branching fractions have been measured for the first time:
$ {\cal{B}}(B^+\to K^+\pi^-\pi^+) = (55.6\pm5.8\pm7.7)\times10^{-6}$ 
and 
$ {\cal{B}}(B^+\to K^+K^-K^+)     = (35.3\pm3.7\pm4.5)\times10^{-6}$.
We present the first observation of the decay $B^+\to f_0(980)K^+$ with 
a branching fraction product of
${\cal{B}}(B^+\to f_0(980)K^+)\times{\cal{B}}(f_0(980)\to \pi^+\pi^-)=
(9.6^{+2.5+1.5+3.4}_{-2.3-1.5-0.8})\times10^{-6}$.
This is the first reported example of a $B$ meson decay to a scalar
pseudoscalar final state. We also report the first observation of
$B^+\to K^*(892)^0\pi^+$ decay with a branching fraction of
${\cal{B}}(B^+\to K^*(892)^0\pi^+) =
(19.4^{+4.2+2.1+3.5}_{-3.9-2.1-6.8})\times10^{-6}$.

\end{abstract}
\pacs{13.20.He, 13.25.Hw, 13.30.Eg, 14.40.Nd}  

\maketitle
%{\renewcommand{\thefootnote}{\fnsymbol{footnote}}}
%\setcounter{footnote}{0}

%\onecolumn
%\twocolumn
%\normalsize

\section{Introduction}

   During the last few  years, a considerable amount of new information
on charmless hadronic decays of $B$ mesons has been reported, primarily
by the CLEO Collaboration.  The discoveries of the $B \to K\pi$ and 
$B \to \pi\pi$ decay modes~\cite{cleokpi} have provided a real basis for 
searches for direct CP-violating effects in the $B$ meson system.

   However, because of large combinatoric backgrounds, studies of 
charmless $B$ decays have concentrated mainly on two-body decay processes.
Three-body decays could significantly broaden the study of $B$ meson 
decay mechanisms and provide additional possibilities for direct CP
violation searches. In this paper, we report the results of a study of 
charged $B$ meson decays to three charged particle final states $K\pi\pi$,
$KK\pi$, and $KKK$, where no  assumptions are made about intermediate 
hadronic resonances. We also present the results of a study of 
quasi-two-body intermediate states in the $K^+\pi^+\pi^-$ and $K^+K^+K^-$
final states. The inclusion of charge conjugate states is implicit 
throughout this work.

   The data sample used for this analysis was collected with the Belle 
detector operating at the KEKB asymmetric energy $e^+e^-$ 
collider~\cite{KEKB}. It consists of 29.1~fb$^{-1}$ taken at the 
$\Upsilon(4S)$ resonance, corresponding to $31.3\times10^{6}$ produced 
$B\bar{B}$ pairs, and 2.3~fb$^{-1}$ taken 40~MeV below the $B\bar{B}$ 
production threshold to perform systematic studies of the 
$e^+e^-\to q\bar{q}$ background.

%%%%%%%%%%%%%%%%%%%%%%%%%%%%%%%%%%%%%%%%%%%%%%%%%%%%%%%%%%%%%%%%%%%%%%
%%%%%%%%%%%%%%%%%%%%%  BELLE  DETECTOR  %%%%%%%%%%%%%%%%%%%%%%%%%%%%%%
%%%%%%%%%%%%%%%%%%%%%%%%%%%%%%%%%%%%%%%%%%%%%%%%%%%%%%%%%%%%%%%%%%%%%%

\section{The Belle detector}

  The Belle detector~\cite{NIM} is a large-solid-angle spectrometer
based on a 1.5~T superconducting solenoid magnet. Charged particle 
tracking is provided by a three layer double-sided silicon vertex 
detector (SVD) and a 50 layer cylindrical drift chamber (CDC) that 
surround the interaction region. The charged particle acceptance 
covers the laboratory polar angle between $\theta=17^{\circ}$ and 
$150^{\circ}$, corresponding to about 92\% of the full solid angle
in the center-of-mass (c.m.) frame. The momentum resolution  is 
determined from  cosmic rays and $e^+ e^-\to\mu^+\mu^-$ events to be 
$\sigma_{p_t}/p_t = (0.30 \oplus 0.19 p_t)\%$, where $p_t$ is the 
transverse momentum in GeV/$c$.

  Charged hadron identification is provided by $dE/dx$ measurements 
in the CDC, an array of 1188 aerogel \v{C}erenkov counters (ACC), and 
a barrel-like array of 128 time-of-flight scintillation counters (TOF).
At large momenta (\mbox{$>2.5$}~GeV/$c$) only the ACC and $dE/dx$ are used 
for separation of charged pions and kaons since here the TOF provides 
no additional discrimination.

   Electromagnetic showering particles are detected in an array of 
8736 CsI(Tl) crystals that is located in the magnetic volume and 
covers the same solid angle as the charged particle tracking system. 
The energy resolution for electromagnetic showers is 
$\sigma_E/E = (1.3 \oplus 0.07/E \oplus 0.8/E^{1/4})\%$, where $E$ is
in GeV. Electron identification in Belle is based on a combination 
of $dE/dx$ measurements in the CDC, the response of the ACC, and the 
position, shape and total energy deposition (i.e. $E/p$) of the shower
registered in the calorimeter. The electron identification efficiency 
is greater than 92\% for tracks with $p_{\rm lab}>1.0$~GeV/$c$ and the 
hadron misidentification probability is below 0.3\%.

   The magnetic field is returned via an iron yoke that is instrumented
to detect muons and $K_L$ mesons. We use a Monte Carlo simulation to 
model the response of the detector and determine acceptance~\cite{GEANT}.

%%%%%%%%%%%%%%%%%%%%%%%%%%%%%%%%%%%%%%%%%%%%%%%%%%%%%%%%%%%%%%%%%%%%%%
%%%%%%%%%%%%%%%%%%%%%  EVENT SELECTION  %%%%%%%%%%%%%%%%%%%%%%%%%%%%%%
%%%%%%%%%%%%%%%%%%%%%%%%%%%%%%%%%%%%%%%%%%%%%%%%%%%%%%%%%%%%%%%%%%%%%%

\section{Event selection}

  Charged tracks are required to satisfy a set of track quality 
requirements based on the average hit residual and on the distances 
of closest approach to the interaction point in the plane 
perpendicular to the beam and the plane containing the beam and the 
track. We also require that the transverse track momenta be greater 
than 0.1~GeV/$c$ to reduce the low momentum combinatoric background.

  Charged kaon candidate tracks are selected with a set of PID criteria 
that has about 90\% efficiency, a charged pion misidentification 
probability of about 8\%, and a negligible contamination from protons.
We also reject tracks that are identified as electrons.
Since the muon identification efficiency and fake rate vary significantly
with the track momentum, we do not reject muons to avoid additional 
systematic error.

  We reconstruct $B$ mesons in three charged track final states with at 
least one positively identified kaon. The candidate events are identified 
by their c.m.\ energy difference, $\Delta E=(\sum_iE_i)-E_{\rm b}$, and 
the beam constrained mass, 
$M_{\rm bc}=\sqrt{E^2_{\rm b}-(\sum_i\vec{p}_i)^2}$, where 
$E_{\rm b}=\sqrt{s}/2$ is the beam energy in the c.m.\ frame, and 
$\vec{p}_i$ and $E_i$ are the c.m.\ three-momenta and energies of the 
candidate $B$ meson decay products. We select events with 
$M_{\rm bc}>5.20$~GeV/$c^2$ and $|\Delta E|<0.20$~GeV, and define 
a {\it signal} region of $|M_{\rm bc}-M_B|<9$~MeV/$c^2$ and 
$|\Delta E|<0.04$~GeV and two $\Delta E$ {\it sideband} regions 
defined as $-0.08$~GeV $<\Delta E<-0.05$~GeV and 
$0.05$~GeV $<\Delta E<0.15$~GeV~\cite{dEreg}. The selection of sideband
regions was based on a Monte Carlo study and was done in such a way that 
the relative fraction of $B\bar{B}$ and $q\bar{q}$ events match that of 
the signal region.
For the normalization factor between the sideband and signal data
samples we use the ratio of areas, namely 0.62.
%The normalization factor between 
%$\Delta E$ sideband and signal regions is then found to be 0.62.

  To evaluate signal and background levels, we require that  one of 
$\Delta E$ or $M_{\rm bc}$ fall in its signal region and examine the 
distribution of candidates in the other, fitting to the sum of a signal 
distribution and an empirical background.
   The signal shapes in $M_{\rm bc}$ and $\Delta E$ distributions are 
parameterized by a Gaussian and sum of two Gaussians with the same mean, 
respectively. The width of the $M_{\rm bc}$ distribution is primarily due 
to the c.m.\ energy spread and is expected to be the same for each channel;
in the fit we fix it at the value $\sigma_{M_{\rm bc}}=2.9$~MeV/$c^2$
determined from the $B^+\to\bar{D^0}\pi^+$, $\bar{D^0}\to K^+\pi^-$ events 
in the same data sample. The $\Delta E$ shape for the signal is also 
determined from the $B^+\to \bar{D^0}\pi^+$ events.
For the $M_{\rm bc}$ projection, we parameterize the background with the 
empirical function $f(M_{\rm bc})\propto\sqrt{1-x^2}\exp[-\xi(1-x^2)]$, 
where $x = M_{\rm bc}/E_{\rm b}$ and $\xi$ is a parameter~\cite{ArgusF}.
We fix the $\xi$ value from a study of data below the $B\bar{B}$ production
threshold. We represent the $\Delta E$ background shape with a linear 
function and restrict the fit to the range 
$-0.1$~GeV~$<\Delta E<0.2$~GeV~\cite{dEreg}. 

%%%%%%%%%%%%%%%%%%%%%%%%%%%%%%%%%%%%%%%%%%%%%%%%%%%%%%%%%%%%%%%%%%%%%%
%%%%%%%%%%%%%%%%%%%%%  BACKGROUND SUPPRESSION  %%%%%%%%%%%%%%%%%%%%%%%
%%%%%%%%%%%%%%%%%%%%%%%%%%%%%%%%%%%%%%%%%%%%%%%%%%%%%%%%%%%%%%%%%%%%%%

\section{Background suppression}

   An important issue for this analysis is the suppression of the 
large combinatoric background which is dominated by $e^+e^-\to~q\bar{q}$
continuum events. We suppress this background with variables that 
characterize the event topology.

   Since the two $B$ mesons produced from $\Upsilon (4S)$ decay 
are nearly at rest in the c.m.\ frame, the angles of the decay products
of the two $B$'s are uncorrelated and the events tend to be spherical. 
In contrast, hadrons from continuum $q\bar{q}$ events tend to exhibit 
a two-jet structure. We use $\theta_{\rm thr}$, which is the angle 
between the thrust axis of the $B$ candidate and that of the rest of 
the event to discriminate between the two cases. The distribution of 
$|\cos\theta_{\rm thr}|$ is strongly peaked near 
$|\cos\theta_{\rm thr}|=1.0$ for $q\bar{q}$ events and is nearly flat 
for $B\bar{B}$ events. We require $|\cos\theta_{\rm thr}|<0.80$ for 
all three-body final states; this eliminates 83\% of the continuum 
background and retains 79\% of the signal events. 

   After imposing the $\cos\theta_{\rm thr}$ requirement, the 
remaining $q\bar{q}$ and $B\bar{B}$ events still have some differences
in topology that are exploited for further continuum suppression. 
We divide the space around the $B$ candidate thrust axis into nine 
polar angle intervals of $10^\circ$ each; the $i$-th interval covers
angles from \mbox{($i$-1)$\times 10^\circ$} to \mbox{$i \times 10^\circ$}.
We define the momentum flows, $x_i (i = 1,9)$, into the $i$-th interval
as a scalar sum of the momenta of all charged tracks and neutral showers
directed in that interval. The momentum flows in corresponding forward
and backward intervals are combined~\cite{VCal}.

  Angular momentum conservation provides some additional discrimination
between $B\bar{B}$ and continuum $q\bar{q}$ events. In $q\bar{q}$ 
production, the direction of the candidate thrust axis with respect to 
the beam axis in the c.m.\ frame, $\theta_{T}$, tends to reproduce the
$1 + \cos^2\theta_{T}$ distribution of the primary quarks. The direction
of the $B$ candidate thrust axis for $B\bar{B}$ events is uniform. 
The $B$ candidate direction with respect to the beam axis, $\theta_{B}$,
exhibits a $\sin^2\theta_{B}$ distribution for $B\bar{B}$ events and is 
uniform for $q\bar{q}$ events.

  A Fisher discriminant~\cite{Fisher} is formed from 11 variables: the
nine momentum flow variables, $|\cos\theta_{T}|$, and $|\cos\theta_{B}|$.
The discriminant, $\cal{F}$, is the linear combination
\[ {\cal{F}} = \sum _{i=1}^{11}\alpha_i x_i \]
of the input variables, $x_i$, that maximizes the separation between 
signal and background. The coefficients $\alpha_i$ are determined from
Monte Carlo simulation using a large set of continuum events and 
signal events modeled as $B^+ \to K^+\pi^+\pi^-$. We use the same set 
of coefficients $\alpha_i$ for all three-body final states. 
The separation between the mean values of the signal and background 
distributions is approximately 1.3 times the signal width.

  For the $K\pi\pi$ and $KK\pi$ final states, we impose a requirement
on the Fisher discriminant variable ${\cal{F}}$ that rejects 90\% of 
the remaining continuum background with about 54\% efficiency for the
signal. For the $KKK$ final state, the continuum background is 
much smaller and we make a looser requirement that rejects 53\% of 
continuum background with 89\% efficiency for the signal.

  To determine the dominant sources of background from other decay 
modes of $B$ mesons, we use a large set of Monte Carlo generated 
$B\bar{B}$ events where both $B$ mesons decay generically~\cite{GEANT}.
Most of the $B\bar{B}$ related background is found to originate 
from $B^+\to\bar{D}^0\pi^+$, $B^+\to J/\psi K^+$ and $B^+\to\psi(2S)K^+$
decays. To suppress this type of background we apply the requirements on 
the invariant masses of the two-particle combinations that are described 
below. The background from $B$ semileptonic decays is additionally 
suppressed by the electron veto requirement. The most significant 
background to the $K^+\pi^+\pi^-$ final state from rare $B$ decays 
is found to originate from $B^+\to \eta'K^+$ followed by 
$\eta'\to \pi^+\pi^-\gamma$. We expect about 3\% of these events to satisfy
all the selection criteria. We find no significant background to the 
$K^+K^+K^-$ final state from other known rare decays of $B$ mesons.

%%%%%%%%%%%%%%%%%%%%%%%%%%%%%%%%%%%%%%%%%%%%%%%%%%%%%%%%%%%%%%%%%%%%%%
%%%%%%%%%%%%%%%%%%  RESULTS OF THE ANALYSIS  %%%%%%%%%%%%%%%%%%%%%%%%%
%%%%%%%%%%%%%%%%%%%%%%%%%%%%%%%%%%%%%%%%%%%%%%%%%%%%%%%%%%%%%%%%%%%%%%

\section{Results of the Analysis }

\subsection{$B^+\to K^+\pi^+\pi^-$}

\begin{figure}[htb]          % Figure I
  \centering
  \includegraphics[width=8.6cm]{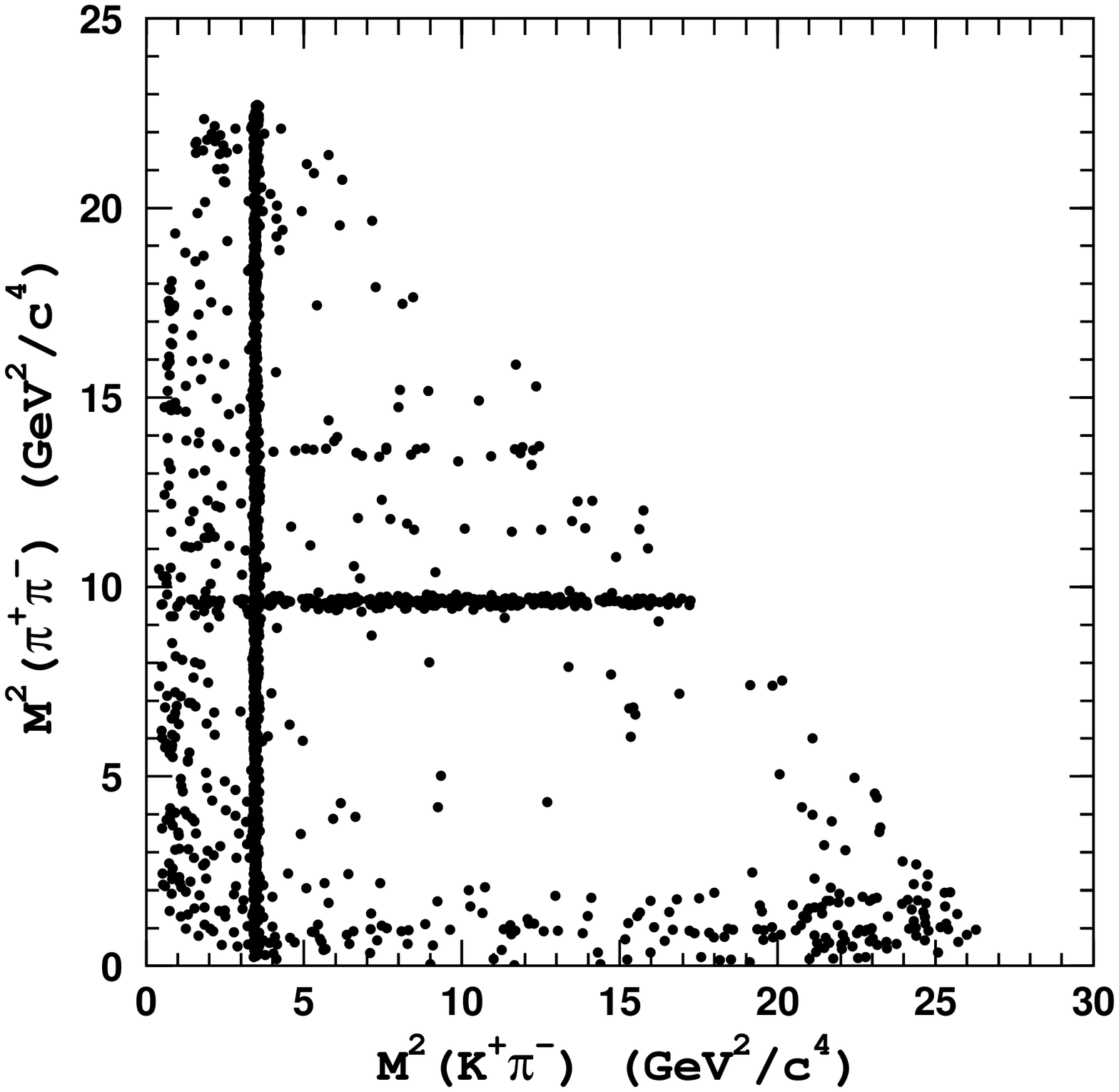}
  \centering
  \caption{The Dalitz plot for $B^+ \to K^+\pi^+\pi^-$ candidates
           from the $B$ signal region.}
  \label{kpp_dp}
\end{figure}

\begin{figure*}[bht]          % Figure II
  \centering
  \includegraphics[width=8.6cm]{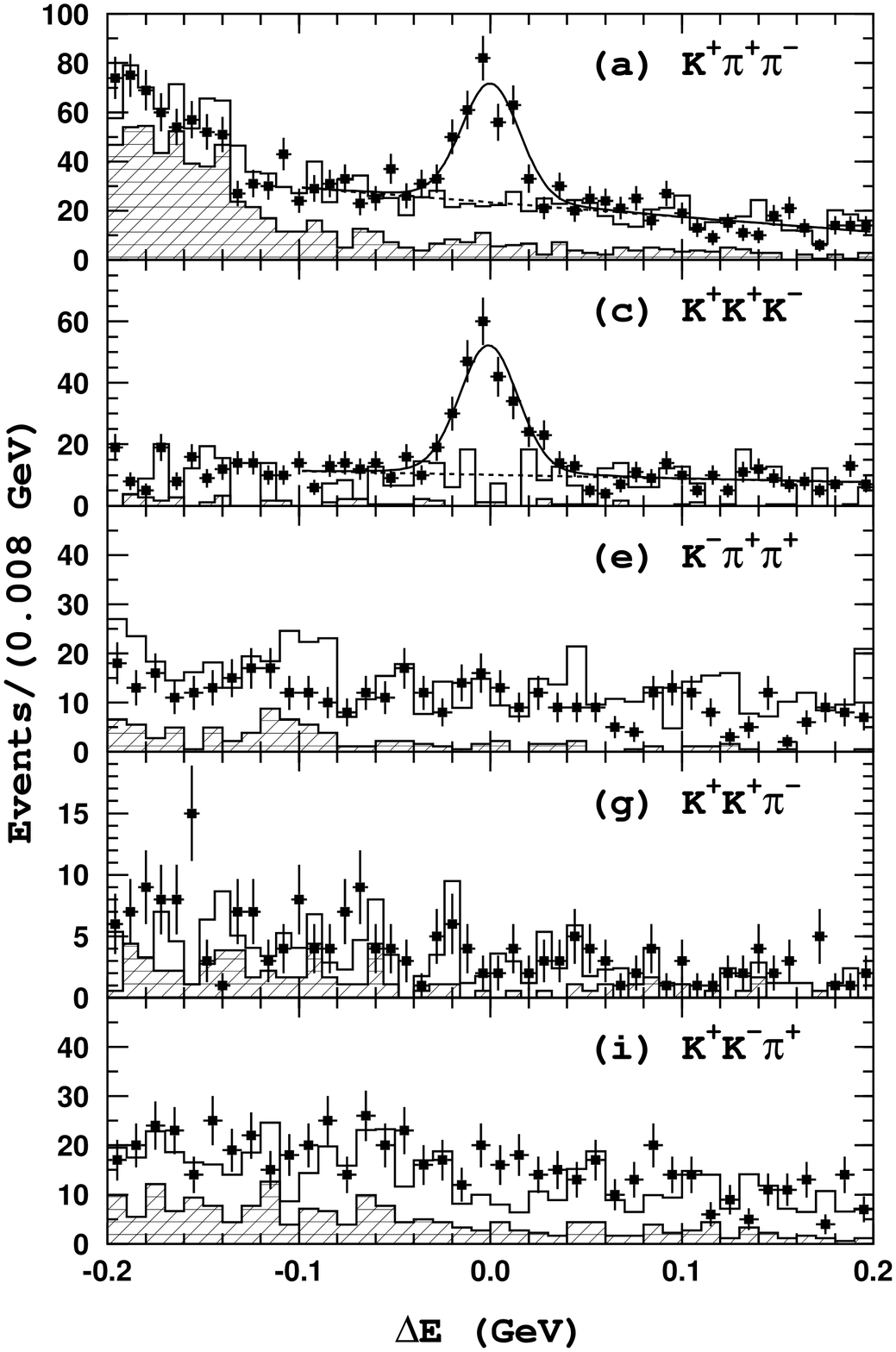} \hfill
  \includegraphics[width=8.6cm]{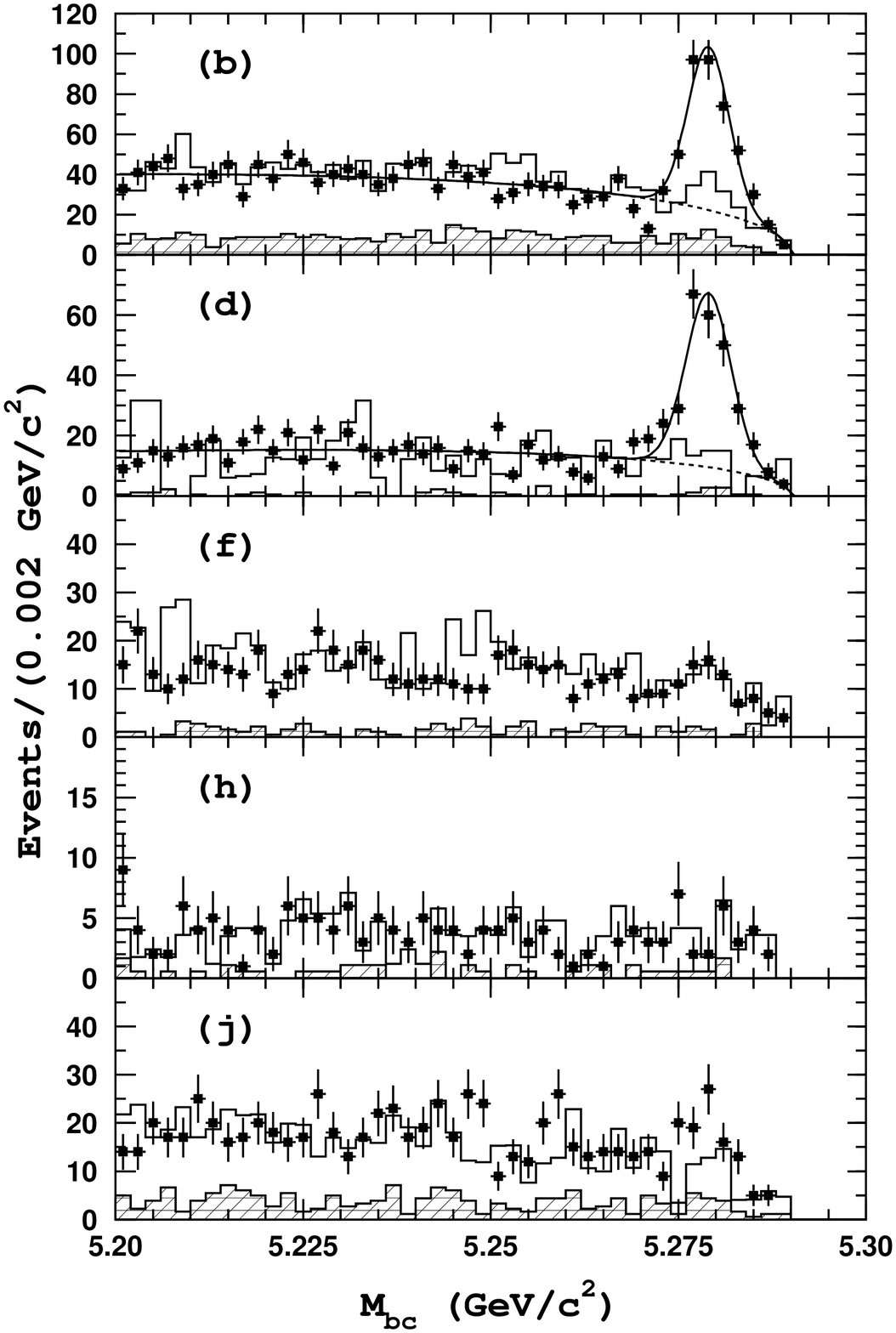}
  \centering
  \caption{The $\Delta E$ (left) and $M_{\rm bc}$ (right) 
           distributions for $B^+\to K^+h^+h^-$ final states: 
           (a,~b)~-~$K^+\pi^+\pi^-$;
           (c,~d)~-~$K^+K^+K^-$;
           (e,~f)~-~$K^-\pi^+\pi^+$;
           (g,~h)~-~$K^+K^+\pi^-$;
           (i,~j)~-~$K^+K^-\pi^+$.
           Points with errors represent data, open histograms are the 
           proper sum of the below-threshold data and $B\bar{B}$ Monte 
           Carlo; the hatched histograms show the contribution of 
           $B\bar{B}$ Monte Carlo only. The solid lines display the 
           signal plus background combined shape.
           The dashed lines correspond to the background shape only.}
  \label{khh_mbde}
\end{figure*}

   For $B^+\to K^+\pi^+\pi^-$ decays, we form $B$ candidates from three
charged tracks where one track is positively identified as a kaon and the 
other two tracks are consistent with a pion hypothesis. Figure~\ref{kpp_dp}
shows the Dalitz plot for selected $B^+\to K^+\pi^+\pi^-$ candidates in the
$B$ signal region. Large contributions from the $B^+ \to \bar{D}^0\pi^+$,
$\bar{D}^0\to K^+\pi^-$ and $B^+\to J/\psi(\psi(2S))K^+$,
$J/\psi(\psi(2S))\to\mu^+\mu^-$ are apparent in the Dalitz plot. The
$J/\psi(\psi (2S))$ modes contribute to this final state due to muon-pion 
misidentification; the contribution from the $J/\psi(\psi (2S))\to e^+e^-$ 
submode is found to be negligible (less than 0.5\%) after the electron veto
requirement. For further analysis, we exclude $\bar{D}^0$ and 
$J/\psi (\psi(2S))$ signals by imposing requirements on the invariant masses
of two intermediate particles: 
$|M(K^+\pi^-)-M_D|>0.10$~GeV/$c^2$;
$|M(h^+h^-)-M_{J/\psi}|>0.07$~GeV/$c^2$;
$|M(h^+h^-)-M_{\psi(2S)}|>0.05$~GeV/$c^2$,
where $h^+$ and $h^-$ are pion candidates. For the $J/\psi(\psi(2S))$
rejection, we use the muon mass hypothesis for charged tracks to 
calculate $M(h^+h^-)$. To suppress the background caused by $\pi/K$ 
misidentification, we exclude candidates if the invariant mass of any
pair of oppositely charged tracks from the $B$ candidate is consistent
with the $D\to K\pi$ hypothesis within 15~MeV/$c^2$ ($\sim 2.5\sigma$), 
independently of the particle identification information ($D$ veto).
The $\Delta E$ and $M_{\rm bc}$ distributions for the remaining events are 
presented in Figs.~\ref{khh_mbde}(a) and~\ref{khh_mbde}(b), respectively. 
Here a significant enhancement in the $B$ signal region is observed;
the result of a fit to the $\Delta E$ distribution is presented in 
Table~\ref{results1}. The expected $\Delta E$ and $M_{\rm bc}$ background
distributions, which are the sum of luminosity-scaled below-threshold 
data and generic $B\bar{B}$ Monte Carlo events, are shown as open 
histograms in Figs.~\ref{khh_mbde}(a) and~\ref{khh_mbde}(b); 
the contributions from the $B\bar{B}$  Monte Carlo sample are shown
as hatched histograms. There are no three-body charmless decays included
in the generic $B\bar{B}$ Monte Carlo.

\begin{figure}[t]          % Figure III
  \centering
  \includegraphics[width=8.6cm]{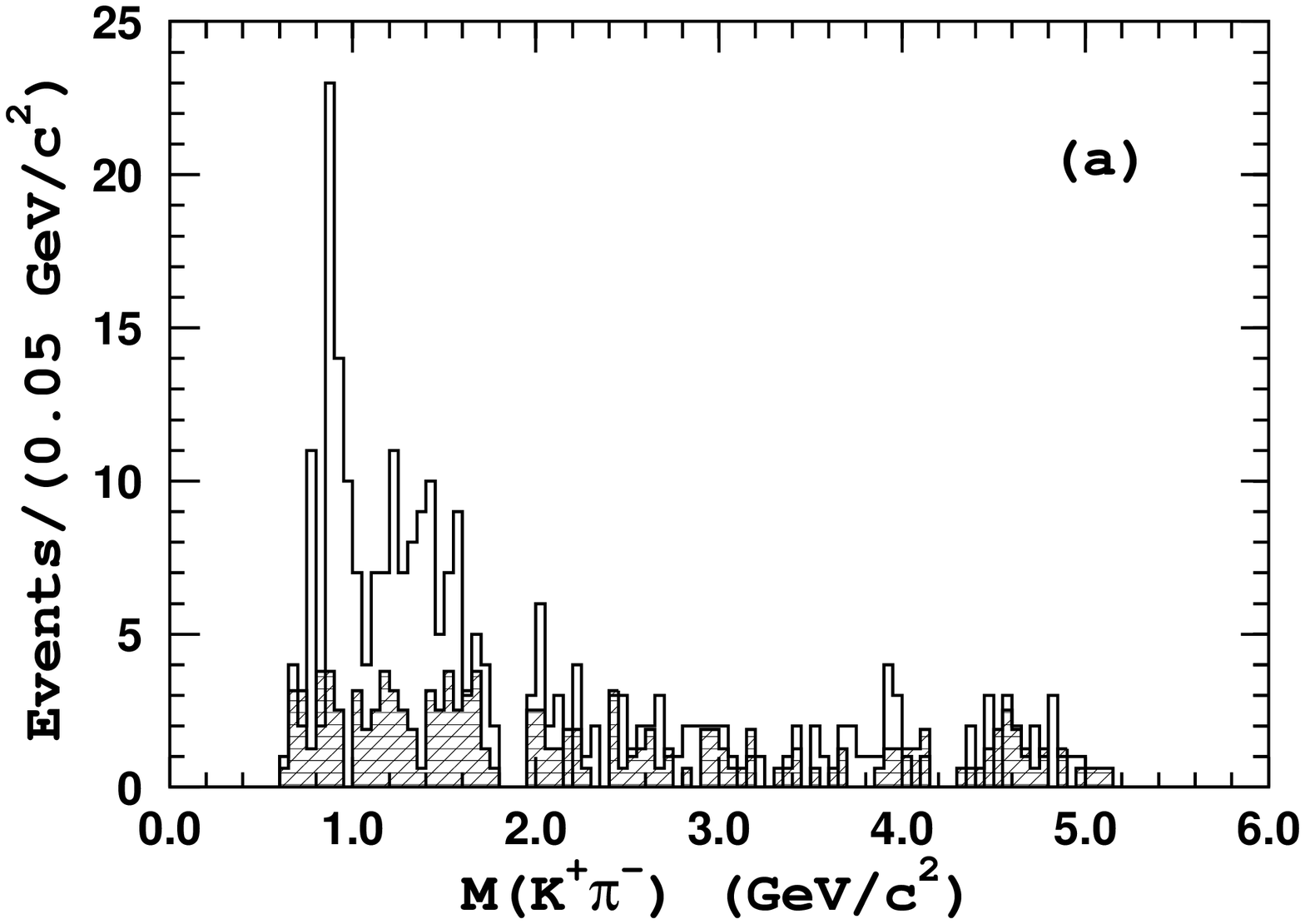} \hfill
  \includegraphics[width=8.6cm]{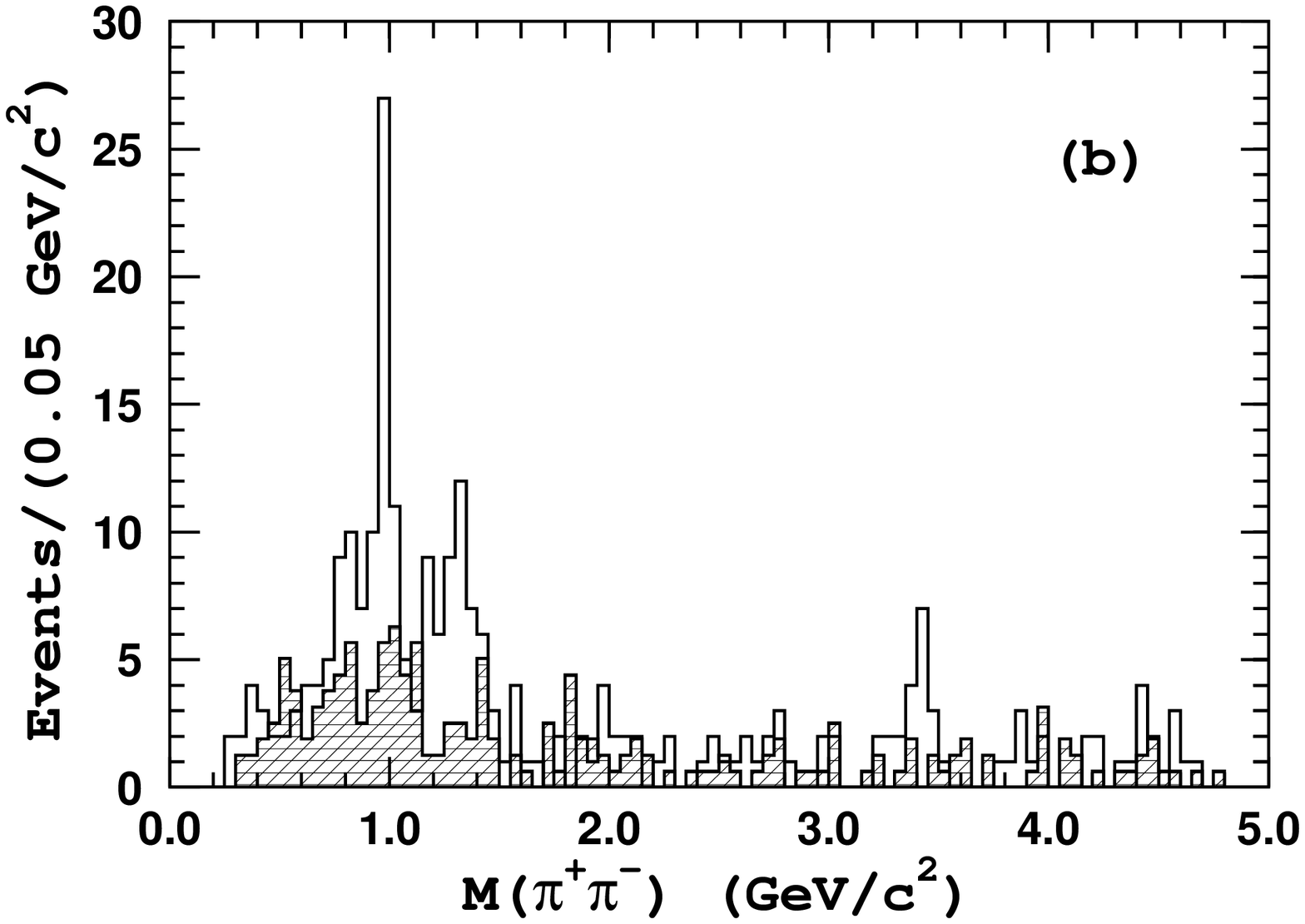}
  \centering
  \caption{ The (a) $K^+\pi^-$ and (b) $\pi^+\pi^-$ invariant mass 
            spectra for selected $B^+\to K^+\pi^+\pi^-$ candidates in 
            the $B$ signal region (open histograms) and for background 
            events in the $\Delta E$ sidebands (hatched histograms).}
  \label{kpp_hhmass}
\end{figure}

   To examine possible intermediate two-body states in the observed 
$B^+\to K^+\pi^+\pi^-$ signal, we analyze the $K^+\pi^-$ and 
$\pi^+\pi^-$ invariant mass spectra shown in Fig.~\ref{kpp_hhmass}.
To suppress the feed-across between the $\pi^+\pi^-$ and $K^+\pi^-$ 
states we require the $K^+\pi^-$ ($\pi^+\pi^-$) invariant mass to be 
larger than 2.0 (1.5)~GeV/$c^2$ when making the $\pi^+\pi^-$ 
($K^+\pi^-$) projection. The hatched histograms shown in 
Fig.~\ref{kpp_hhmass} are the corresponding two-particle invariant 
mass spectra for the background events in the $\Delta E$ sidebands
plotted with a weight of 0.62.

   The $K^+\pi^-$ invariant mass spectrum is characterized by a 
narrow peak  around 0.9~GeV/$c^2$ which is identified as the $K^*(892)^0$
and a broad enhancement above 1.0~GeV/$c^2$ which is subsequently 
referred to as $K_X(1400)$. In the $\pi^+\pi^-$ invariant mass 
spectrum two distinct structures in the low mass region are observed.
One is slightly below 1.0~GeV/$c^2$ and is identified as the $f_0(980)$ 
while the other is between 1.0~GeV/$c^2$ and 1.5~GeV/$c^2$ and is
referred to as $f_X(1300)$. Some excess of signal events can be also
observed in the $\rho^0(770)$ mass region. The peak around 
3.4~GeV/$c^2$ is consistent with the process $B^+\to\chi_{c0}K^+$, 
$\chi_{c0}\to\pi^+\pi^-$, and is the subject of a separate 
analysis~\cite{chi_c0}. In this paper we exclude the 
$B^+\to\chi_{c0}K^+$ candidates from the analysis of two-body final 
states by applying the requirement on the $\pi^+\pi^-$ invariant mass: 
$|M(\pi^+\pi^-)-M_{\chi_{c0}}|>0.05$~GeV/$c^2$.

   For further analysis we subdivide the full Dalitz plot area into 
seven non-overlapping regions as defined in Table~\ref{kpp_fits}.
Regions from I to V are arranged to contain the major part of the 
signal from the $B^+\to K^*(892)^0\pi^+$, $B^+\to K_X(1400)\pi^+$,
$B^+\to \rho^0(770)K^+$, $B^+\to f_0(980)K^+$, and 
$B^+\to f_X(1300)K^+$ final states, respectively. The area in the 
Dalitz plot where $K\pi$ and $\pi\pi$ resonances overlap is covered 
by region VI, and region VII covers the rest of the Dalitz plot.
The results of the fits to the $\Delta E$ distributions for all seven
regions are summarized in Table~\ref{kpp_fits}.
The procedure used for the extraction of the two-body branching 
fractions is described in detail in Section~VI.

%%%%%%%%%%%%%%%%%%%%%%%%%%%%%%%%%%%%%%%%%%%%%%%%%%%%%%%%%%%%%%%%%%%%%%%

\subsection{$B^+\to K^+K^+K^-$}

\begin{figure}[htb]          % Figure IV
  \centering
  \includegraphics[width=8.6cm]{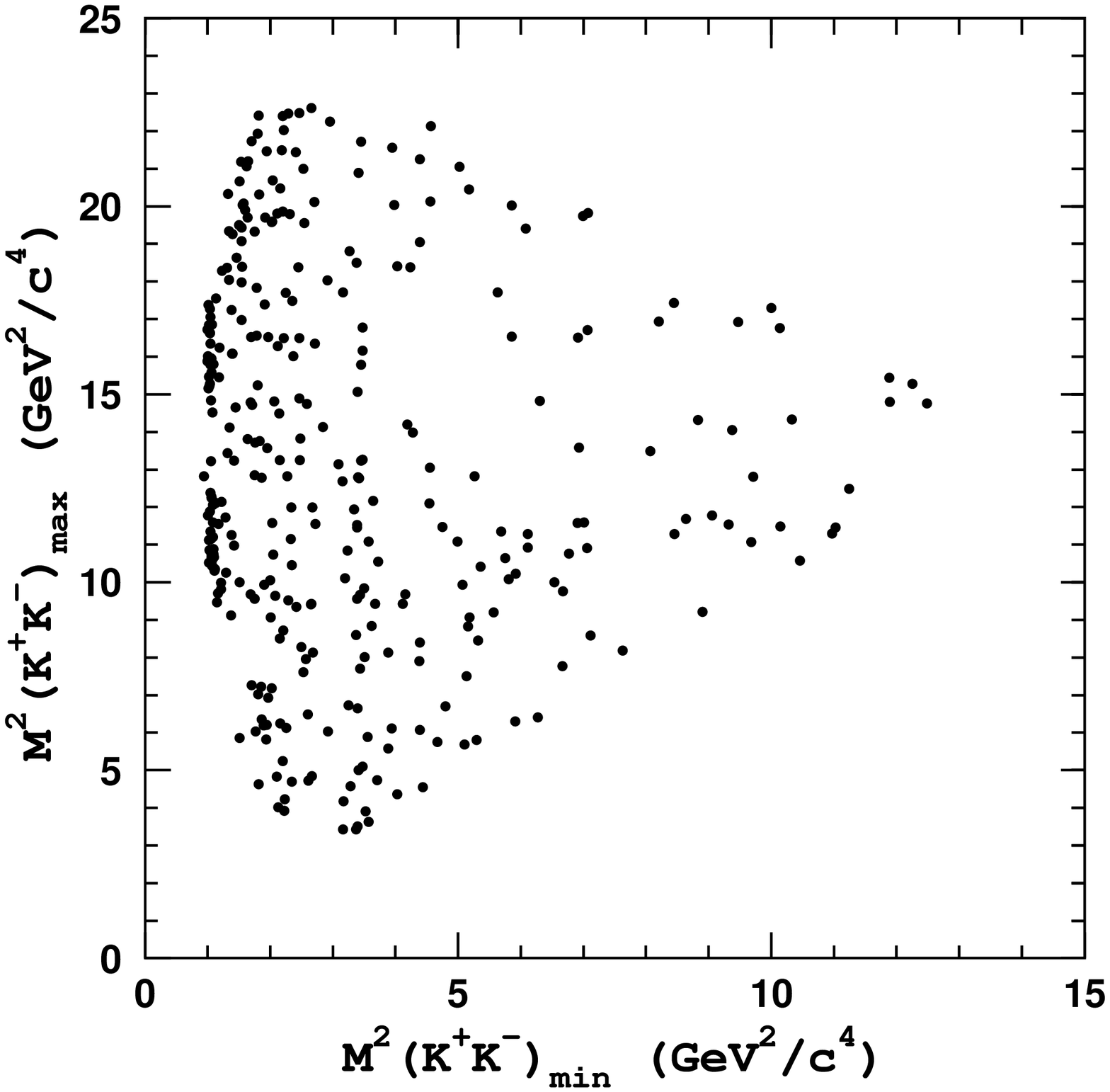}
  \centering
  \caption{The Dalitz plot for $B^+ \to K^+K^+K^-$ candidates
           from the $B$ signal region.}
  \label{kkk_dp}
\end{figure}

   For the selection of $B^+\to K^+ K^+ K^-$ events, we use combinations 
of three charged tracks that are positively identified as kaons. 
The Dalitz plot for selected $B^+\to K^+K^+K^-$ candidate events in the
$B$ signal region after the $D$ veto is shown in Fig.~\ref{kkk_dp}.
Since in this case there are two same-charge kaons, we distinguish 
the $K^+K^-$ combinations with smaller, $M(K^+K^-)_{\rm min}$, and larger,
$M(K^+K^-)_{\rm max}$, invariant masses. We avoid double entries per 
candidate by forming the Dalitz plot as $M^2(K^+K^-)_{\rm max}$ versus 
$M^2(K^+K^-)_{\rm min}$. The signal from the Cabibbo-suppressed 
$B^+\to \bar{D^0}K^+$, $\bar{D^0}\to K^+K^-$ decay mode is apparent as a
vertical strip in the Dalitz plot. The corresponding Cabibbo-allowed 
$B^+\to \bar{D^0}\pi^+$, $\bar{D^0}\to K^+K^-$ decays can also contribute
to this final state as a result of pion-kaon misidentification. We exclude
candidates consistent with the $B^+\to \bar{D^0}h^+$ hypothesis from 
further analysis by imposing the requirement on the $K^+K^-$ invariant 
mass $|M(K^+K^-)-M_{D^0}|>0.025$~GeV$/c^2$. The $\Delta E$ and
$M_{\rm bc}$ distributions after the exclusion of $D$ mesons are presented
in Figs.~\ref{khh_mbde}(c) and~\ref{khh_mbde}(d) respectively. A large 
peak in the $B$ signal region is apparent in both distributions. The 
result of a fit to the $\Delta E$ distribution is presented in 
Table~\ref{results1}.

\begin{figure*}[htb]          % Figure V
  \centering
  \includegraphics[width=8.6cm]{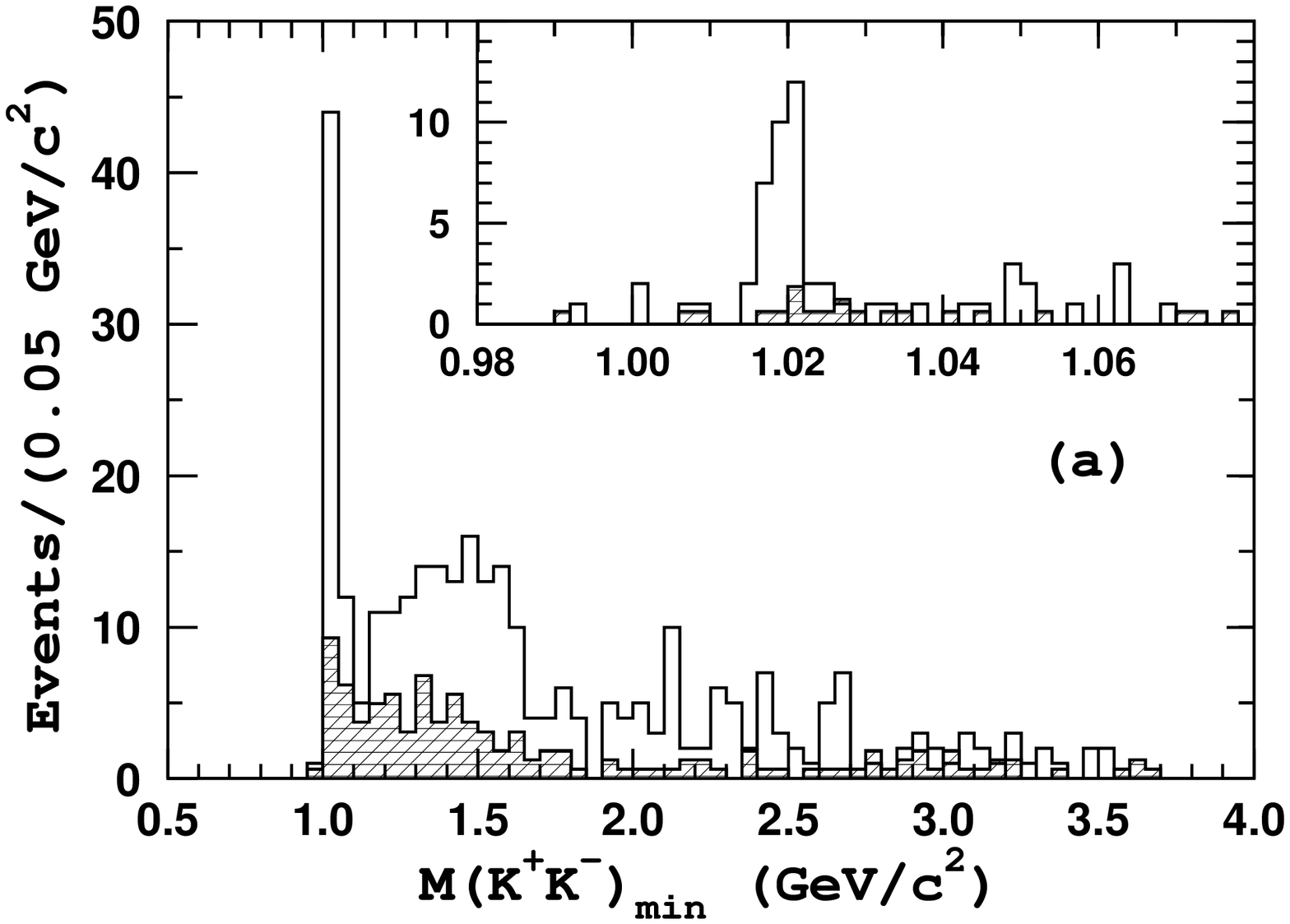} \hfill
  \includegraphics[width=8.6cm]{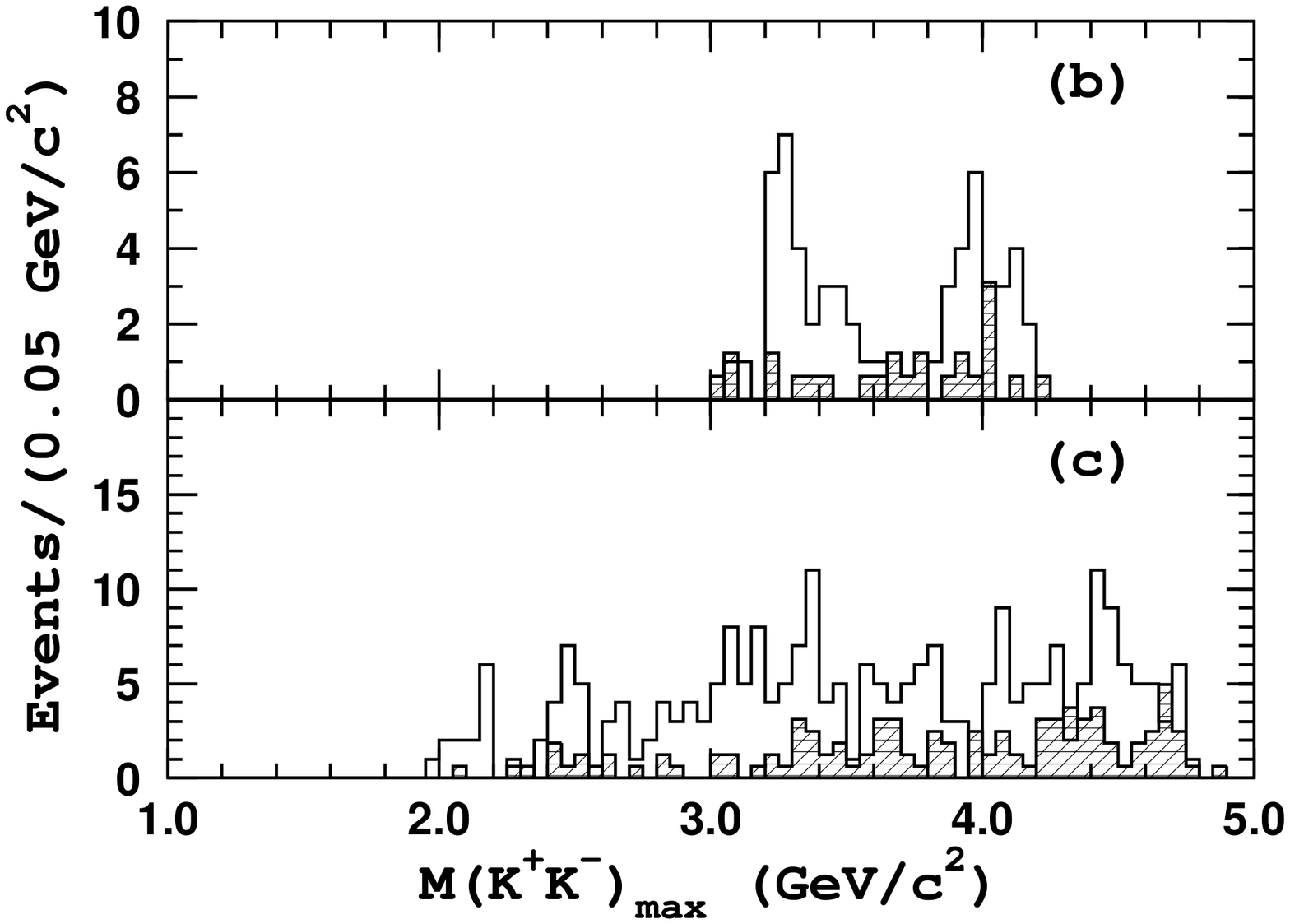}
  \centering
  \caption{ The $K^+K^-$ invariant mass spectra for selected 
            $B^+\to K^+K^+K^-$ candidates in the $B$ signal region 
            (open histograms) and for background events in the 
            $\Delta E$ sidebands (hatched histograms).
            (a)~The $K^+K^-$ combination with the smaller invariant 
                mass. The inset shows the $\phi(1020)$ region in
                2~MeV/c$^2$ bins;
            (b)~the $M(K^+K^-)_{\rm max}$ spectrum with 
                $M(K^+K^-)_{\rm min}<1.1$~GeV/$c^2$ and
            (c)~the $M(K^+K^-)_{\rm max}$ spectrum with
                $M(K^+K^-)_{\rm min}>1.1$~GeV/$c^2$.}
  \label{kkmass}
\end{figure*}

  The $K^+K^-$ invariant mass spectra for events from the $B$ signal region
are shown as open histograms in Figs.~\ref{kkmass}(a)-\ref{kkmass}(c). 
The hatched histograms show the corresponding spectra for background events
in the $\Delta E$ sidebands, plotted with a weight of 0.62. 
The $M(K^+K^-)_{\rm min}$ spectrum, shown in Fig.~\ref{kkmass}(a), is 
characterized by a narrow peak at 1.02~GeV/$c^2$ corresponding to the 
$\phi(1020)$ meson and a broad structure around 1.5~GeV/$c^2$; this is 
subsequently referred to as $f_X(1500)$. To plot the $M(K^+K^-)_{\rm max}$
mass spectrum we subdivide the $M(K^+K^-)_{\rm min}$ mass region into two
ranges: $M(K^+K^-)_{\rm min}<1.1$~GeV/$c^2$ and 
$M(K^+K^-)_{\rm min}>1.1$~GeV/$c^2$. The $M(K^+K^-)_{\rm max}$ mass 
spectra for these two regions are presented in Fig.~\ref{kkmass}(b)
and Fig.~\ref{kkmass}(c) respectively. The prominent structure observed
in Fig.~\ref{kkmass}(b) reflects the 100\% $\phi$ meson polarization in 
the $B^+\to\phi K^+$ decay due to angular momentum conservation. 
In contrast, the distribution of signal events in Fig.~\ref{kkmass}(c) 
is quite uniform after the background is subtracted. For the analysis of 
two-body final states we exclude events that are consistent with the 
$B^+\to\chi_{c0}K^+$, $\chi_{c0}\to K^+K^-$ decay by applying the 
requirement on the $K^+K^-$ invariant mass: 
$|M(K^+K^-)-M_{\chi_{c0}}|>0.05$~GeV/$c^2$.

   For further analysis we subdivide the full Dalitz plot area 
into the four non-overlapping regions defined in Table~\ref{kkk_fits}.
Regions I and II are arranged to contain the major part of the signal 
from the $B^+\to \phi(1020)K^+$ and $B^+\to f_X(1500)K^+$ final states,
respectively. Regions III and IV cover the remaining part of the Dalitz 
plot. The results of the fits to the $\Delta E$ distributions for all 
four regions are summarized in Table~\ref{kkk_fits}.

%%%%%%%%%%%%%%%%%%%%%%%%%%%%%%%%%%%%%%%%%%%%%%%%%%%%%%%%%%%%%%%%%%%%%%

\subsection{$B^+\to K^-\pi^+\pi^+$, $B^+\to K^+K^+\pi^-$ and 
$B^+\to K^+K^-\pi^+$}

% and $B^+\to K^+K^-\pi^+$ 
   In general, we do not expect any signal in the $B^+\to K^-\pi^+\pi^+$
and $B^+\to K^+K^+\pi^-$ final states. The Standard Model prediction for 
the $B^+\to K^+K^+\pi^-$ branching fraction is of the order of $10^{-11}$, 
and even much smaller for the $B^+\to K^-\pi^+\pi^+$ final 
state~\cite{b2ssd}. However, these signals could be significantly enhanced 
in some extensions of the Standard Model~\cite{nonsm}, and, thus, these 
modes can be used to search for physics beyond the Standard Model.

\begin{figure}[htb]          % Figure VI
\centering
  \includegraphics[width=8.6cm]{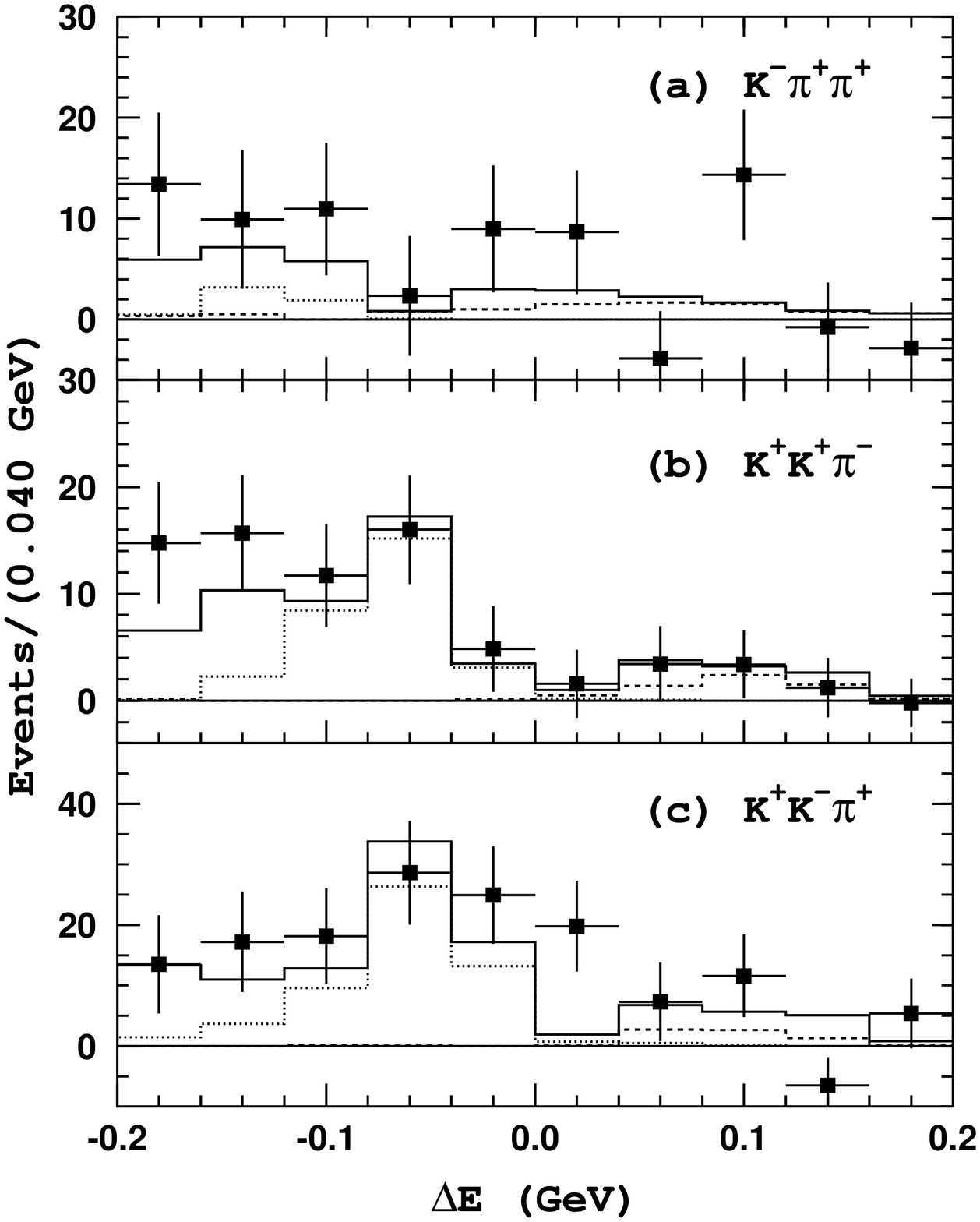}
  \centering
  \caption{Results of the fit to the $M_{\rm bc}$ distributions in
           $\Delta E$ bins.
           The data (points with errors) are compared with the MC 
           expectation (open histogram). The feed-down from 
           $K^+K^+K^-$ and $K^+\pi^+\pi^-$ final states is shown by
           the dotted and dashed histograms, respectively.}
  \label{khh_de_2}
\end{figure}

   For the $K^+K^+\pi^-$ final state, we reject candidates that are 
consistent with the $B^+\to \bar{D^0}K^+$, $\bar{D^0}\to K^+\pi^-$ 
decay by imposing the requirement on the $K^+\pi^-$ invariant mass
$|M(K^+\pi^-)-M_{D^0}|>0.10$~GeV/$c^2$. In case of the $K^+K^-\pi^+$ 
channel we reject candidates that are consistent with the 
$B^+\to \bar{D^0}\pi^+$, $\bar{D^0}\to K^+K^-$ decay with the 
requirement $|M(K^+K^-)-M_{D^0}|>0.05$~GeV/$c^2$. We also apply the $D$ 
veto requirement for the three modes. The resulting $\Delta E$ and 
$M_{\rm bc}$ distributions for the $K^-\pi^+\pi^+$, $K^+K^+\pi^-$ and 
$K^+K^-\pi^+$ final states are presented in 
Figs.~\ref{khh_mbde}(e)-\ref{khh_mbde}(j). Although we do not observe any 
signal in the $\Delta E$ distributions, there is an excess of events in 
the signal region of the $M_{\rm bc}$ distributions for the $K^-\pi^+\pi^+$ 
and $K^+K^-\pi^+$ final states. These excesses could be caused by 
incorrectly reconstructed $B$ decays. To subtract this background we 
subdivide the $\Delta E$ region into ten bins of 40~MeV width and 
determine the signal yield in each bin from the fit to the corresponding 
$M_{\rm bc}$ spectrum. The results of the fit, along with the expected 
contributions from the generic $B\bar{B}$ decays and the feed-down due 
to the particle misidentification from the $B^+\to K^+\pi^+\pi^-$ and 
$B^+\to K^+K^+K^-$ decay modes, are presented in Fig.~\ref{khh_de_2}. 
The latter two components are shown in Fig.~\ref{khh_de_2} by the dotted 
and dashed histograms, respectively. The excess of events over the total 
expected background in the $\Delta E$ signal region (two bins around 
$\Delta E = 0$) is considered to be a signal yield. The results are 
summarized in Table~\ref{results1}. We do not observe a statistically
significant signal in any of these three-body modes.

   The feed-across between $K^+\pi^+\pi^-$ and $K^+K^+K^-$ final states
is found to be negligible. True $B^+\to K^+K^+K^-$ events reconstructed
as $K^+\pi^+\pi^-$ contribute mainly to the $\Delta E < -0.10$~GeV region 
that is excluded from the fit. The fraction of true $K^+\pi^+\pi^-$ events
improperly reconstructed as $K^+K^+K^-$ is less than 0.1\%.

%%%%%%%%%%%%%%%%%%%%%%%%%%%%%%%%%%%%%%%%%%%%%%%%%%%%%%%%%%%%%%%%%%%%%%
%%%%%%%%%%%%%%%%%%%%  BRANCHING FRACTIONS RESULTS  %%%%%%%%%%%%%%%%%%%
%%%%%%%%%%%%%%%%%%%%%%%%%%%%%%%%%%%%%%%%%%%%%%%%%%%%%%%%%%%%%%%%%%%%%%

\section{Branching Fractions}

   To determine branching fractions, we normalize our results to the 
observed $B^+\to \bar{D}^0\pi^+$, $\bar{D}^0\to K^+\pi^-$ signal. This 
removes systematic effects in the particle identification efficiency, 
charged track reconstruction efficiency and the systematic uncertainty 
due to the cuts on event shape variables. We calculate the branching 
fraction for $B$ meson decay to a particular final state $f$ via the 
relation
\[ {\cal{B}}(B^+\to f) = 
   {\cal{B}}(B^+\to\bar{D}^0\pi^+){\cal{B}}(\bar{D}^0\to K^+\pi^-)
   \frac{N_f}{N_{D\pi}}\frac{\varepsilon_{D\pi}}{\varepsilon_{f}},
\]
where $N_f$ and $N_{D\pi}$ are the numbers of reconstructed events for 
the final state $f$ and that for the reference process, respectively;
$\varepsilon_{f}$ and $\varepsilon_{D\pi}$ are the corresponding 
reconstruction efficiencies. We use the signal yield extracted from the 
fit to the corresponding $\Delta E$ distribution; we do not use the 
$M_{\rm bc}$ distribution, because it, in general, suffers more from 
the $B\bar{B}$  background.

   The number of signal events for the reference process
$B^+\to \bar{D}^0\pi^+$, $\bar{D}^0\to K^+\pi^-$ is found to be 
$1349\pm40$ for the $K^+K^+K^-$ selection requirements and $805\pm32$
for the requirements used for all other three-body combinations. 
The corresponding reconstruction efficiencies are $26.8$\% and $16.1$\%, 
respectively. The reconstruction efficiency for each three-body final 
state is determined from the Monte Carlo simulation of events that are 
generated with a uniform Dalitz plot distribution. The branching 
fraction results for $K^+\pi^+\pi^-$ and $K^+K^+K^-$ final states are 
presented in Table~\ref{results1}, where the first quoted error is 
statistical and the second is systematic. The dominant sources of systematic
error are listed in Table~\ref{khh_syst}.
We estimate the systematic uncertainty due to variations of reconstruction 
efficiency over the Dalitz plot using two sets of MC data generated with
uniform distribution (phase space) and using some model (described below). 
The uncertainty due to the particle identification is estimated using pure 
samples of kaons and pions from $D^0\to K^+\pi^-$ decays, where the
$D^0$ flavor is tagged using $D^{*+}\to D^0\pi^+$ decays.
To estimate the uncertainty due to the signal and background shapes
parameterization, we fit the $\Delta E$ distributions using different
functions for the background description (linear, parabolic,
exponential plus constant) and varying the parameters of the signal function
(sum of two Gaussians with the same mean) within their errors.

   Since we do not observe a statistically significant signal in the 
$K^-\pi^+\pi^+$, $K^+K^+\pi^-$ or $K^+K^-\pi^+$ final states, we place 
the 90\% confidence level upper limits on their branching fractions.
These limits are given in Table~\ref{results1}. To calculate the upper 
limits, we follow the PDG recommendation~\cite{pdgul}.

%%%%%%%%%%%%%%%%%%%%%%%%%%%%%%%%%%%%%%%%%%%%%%%%%%%%%%%%%%%%%%%%%%%%%%

\subsection{Exclusive two-body branching fractions in the 
$K^+\pi^+\pi^-$ final state}

   In the determination of the branching fractions for exclusive 
two-body final states, we have to take into account the possibility of 
interference between wide resonances. This requires some assumptions 
about the states that are being observed and, as a consequence, 
introduces some model dependence into the extraction of the exclusive 
branching fractions. The present level of statistics does not permit 
unambiguous interpretation of the $K_X(1400)$ and $f_X(1300)$ states and,
thus, it is not possible to use the data to fix all of the input model 
parameters. For this analysis we assume that the observed $K_X(1400)$
and $f_X(1300)$ states are $0^{+}$ scalars. While this does not 
contradict the observed signal, some contributions from vector ($1^{-}$) 
and tensor ($2^{+}$) resonances cannot be excluded. The uncertainty 
related to this assumption is included in the model-dependent error 
described below. We ascribe to the $K_X(1400)$ state the parameters of 
$K^*_0(1430)$ ($M$ = 1412 MeV/$c^2$, $\Gamma$ = 294 MeV) and to 
$f_X(1300)$ state the parameters of $f_0(1370)$ ($M$ = 1370 MeV/$c^2$,
$\Gamma$ = 400 MeV)~\cite{PDG}.

   For further analysis we make the following assumptions:
\begin{itemize}
  \item{ The observed signal in the $K^+\pi^+\pi^-$ final state can be 
         described by some number of two-body final states. 
         We restrict ourselves to the following set of exclusive 
         two-body final states: $K^*(892)^0\pi^+$, $K_X(1400)\pi^+$,
         $\rho^0(770)K^+$, $f_0(980)K^+$ and $f_X(1300)K^+$. 
         We enumerate these final states as 1 through 5 in the order 
         mentioned above.}

  \item{ Given this set of two-body final states, we determine the 
         exclusive branching fractions neglecting the effects of 
         interference. The uncertainty due to possible interference
         between different intermediate states is included 
         in the final result as a model-dependent error.}
\end{itemize}

   In order to extract the signal yield for each two-body final state, 
we perform a simultaneous likelihood fit to the $\Delta E$ distributions
for the seven regions of the $K^+\pi^+\pi^-$ Dalitz plot. We express the 
expected number $n_j$ of signal events in the $j$-th region of the Dalitz 
plot as a linear combination
\[ n_j = \sum^{5}_{i=1}\varepsilon_{ij}N_i,\]
where $N_i$ is the total number of signal events in the $i$-th two-body 
final state and $\varepsilon_{ij}$ is the probability for the $i$-th 
final state to contribute to the $j$-th region in the Dalitz plot. 
The $\varepsilon_{ij}$ matrix is determined from the Monte Carlo 
simulation and includes the reconstruction efficiency. This procedure 
takes into account the effect of correlations between different channels 
in the determination of the statistical errors.

   The results of the fit are summarized in Table~\ref{kpp_min}.
Combining all the relevant numbers, we calculate the product of branching 
fractions ${\cal{B}}(B^+\to Rh^+)\times{\cal{B}}(R\to h^+h^-)$, where 
$R$ denotes the two-body intermediate resonant state. We present three 
types of errors for the branching fractions: the first error is 
statistical, the second is systematic, and the third reflects the 
model-dependent uncertainty. In general, the model-dependent error is 
due to uncertainties in the effects of interference between different 
resonant states. We estimate this error by means of a 
$B^+\to K^+\pi^+\pi^-$ Monte Carlo simulation that includes interference 
effects between all the final states mentioned above. We vary the 
relative phases of the resonances and determine the signal yield using the
procedure described above. The maximal deviations from the central 
values are used as an estimate of the model dependence of the obtained 
branching fractions.
   We find that the model-dependent errors associated with the wide 
resonances introduce significant uncertainties into the branching 
fraction determination. In the case of the $\rho^0(770)K^+$ final state,
this effect is enhanced by the smallness of the signal itself. Since we 
do not observe a significant signal in this channel, we report a 90\% 
confidence level upper limit. The statistical significance of the signal,
in terms of the number of standard deviations, is calculated as 
$\sqrt{-2\ln({\cal{L}}_0/{\cal{L}}_{\rm max})}$, where ${\cal{L}}_{\rm max}$
and ${\cal{L}}_{0}$ denote the maximum likelihood with the nominal signal 
yield and with the signal yield fixed at zero, respectively.

%%%%%%%%%%%%%%%%%%%%%%%%%%%%%%%%%%%%%%%%%%%%%%%%%%%%%%%%%%%%%%%%%%%%%%%%%%

\subsection{Exclusive two-body branching fractions in the
$K^+K^+K^-$ final state}

   In the case of the three charged kaon final state, we clearly observe 
the $\phi(1020)$ meson plus a very broad $f_X(1500)$ structure that we 
currently cannot interpret unambiguously. It could be a complex 
superposition of several intermediate states and some contribution from the
non-resonant $B^+\to K^+K^+K^-$ decay is also possible. For our study of 
systematic and model-dependent uncertainties, we construct a simplified 
model and parameterize the $f_X(1500)$ structure as a hypothetical scalar 
state with $M=1500$ MeV/$c^2$ and $\Gamma = 700$ MeV. We find qualitative 
agreement between the experimental Dalitz plot distribution of the signal 
events and that obtained from the Monte Carlo simulation with this simple 
model.

  Then we extract the signal yield for the two-body final states:
$B^+\to\phi(1020)K^+$ and the so-called $B^+\to f_X(1500)K^+$, which
is, in fact, all of the remaining signal. We follow the same procedure 
as we used for the $K^+\pi^+\pi^-$ final state. The signal yields are 
determined from a simultaneous fit to the $\Delta E$ distributions for 
four separate regions of the $K^+K^+K^-$ Dalitz plot.
The results of the fit are summarized in Table~\ref{kkk_min}.

  We determine the model-dependent error in the same way as we did for
the $K^+\pi^+\pi^-$ final state. In the case of the $K^+K^+K^-$ final
state the  model-dependent error is found to be much smaller than
in the $K^+\pi^+\pi^-$ final state. This is mainly due to the small
width of the $\phi(1020)$ meson.

\section{Discussion \& Conclusion}

  The high quality of $\pi/K$ separation at Belle allows us to measure, 
for the first time, the branching ratios for the three-body modes 
$ {\cal{B}}(B^+\to K^+\pi^-\pi^+) = (55.6\pm5.8\pm7.7)\times10^{-6}$ and 
$ {\cal{B}}(B^+\to K^+K^-K^+)     = (35.3\pm3.7\pm4.5)\times10^{-6}$
without assumptions about particular intermediate mechanisms.
CLEO~\cite{berg96b} and BaBar~\cite{babar} have previously placed upper
limits on the branching fractions of non-resonant three-body decays:
${\cal{B}}(B^+ \to K^+\pi^+\pi^-)<28 \times 10^{-6}$ (CLEO),
${\cal{B}}(B^+ \to K^+\pi^+\pi^-)<66 \times 10^{-6}$ (BaBar),
${\cal{B}}(B^+ \to K^+K^+K^-)<38 \times 10^{-6}$ (CLEO).
A comparison of the applied selection 
criteria shows that CLEO and BaBar restricted their analyses to the region 
of invariant mass above 2~GeV/$c^2$ for any pair of the particles. This 
requirement effectively removes most of the low mass resonances that 
provide the dominant contribution to our observed signal. They assume a 
uniform distribution of events over the Dalitz plot to obtain the limits 
quoted above. The upper limits reported here for the $K^-\pi^+\pi^+$, 
$K^+K^+\pi^-$ and $K^+K^-\pi^+$ modes are considerably more restrictive 
than previous limits from CLEO~\cite{berg96b} and OPAL~\cite{OPAL}.

   Significant signals are observed for the first time in the decay modes
$B^+ \to f_0(980) K^+$ and $B^+\to K^*(892)^0\pi^+$. The measured branching
fraction product for the $f_0(980) K^+$ final state is 
${\cal{B}}(B^+\to f_0(980)K^+)\times{\cal{B}}(f_0(980)\to \pi^+\pi^-)=
(9.6^{+2.5+1.5+3.4}_{-2.3-1.5-0.8})\times10^{-6}$. This is the first
observation of a $B$ decay to a charmless scalar-pseudoscalar final state. 
The measured branching fraction product for the $K^*(892)^0\pi^+$ final 
state is 
${\cal{B}}(B^+\to K^*(892)^0\pi^+)\times{\cal{B}}(K^*(892)^0\to K^+\pi^-)=
(12.9^{+2.8+1.4+2.3}_{-2.6-1.4-4.5})\times10^{-6}$. Using the value of 
${\cal{B}}(K^*(892)^0\to K^+\pi^-)=2/3$, we translate our measurement into 
the branching fraction ${\cal{B}}(B^+\to K^*(892)^0\pi^+) =
(19.4^{+4.2+2.1+3.5}_{-3.9-2.1-6.8})\times10^{-6}$. The significant 
enhancement in the $K^+\pi^-$ invariant mass spectrum above the $K^*(892)$ 
mass agrees with the scalar $K_0^*(1430)$ hypothesis. This is also in 
agreement with theoretical predictions~\cite{chernyak} for the 
$B^+\to K_0^*(1430)\pi^+$ branching fraction based on the factorization 
model. Nevertheless, we cannot exclude some contribution from the tensor 
$K_2^*(1430)$ state.

  The interpretation of the peak with a $\pi^+\pi^-$ invariant mass around 
1300 MeV/$c^2$ in the $K^+\pi^+\pi^-$ system is less certain. There are two
known candidate states: the $f_2(1270)$ and $f_0(1370)$~\cite{PDG}. 
Attributing the peak to the $f_0(1370)$, with its rather small coupling to 
$\pi^+\pi^-$~\cite{anisov}, would lead to an unusually large branching 
fraction for a charmless $B$ decay mode. On the other hand, as recently 
shown in~\cite{kim}, the factorization model predicts a very small branching
fraction for the $B^+\to f_2(1270)K^+$. If our observation is, in fact, due
to the $f_2(1270)$, this would provide evidence for a significant 
nonfactorizable contribution. 

  We cannot identify the broad structure observed in the 
$B^+\to K^+K^+K^-$ final state above the $\phi(1020)$ meson.
It is hardly compatible with the presence of a single scalar state,
either $f_0(1370)$ or $f_0(1500)$~\cite{PDG}. We also cannot exclude 
the presence of a non-resonant contribution or the case of several
resonances contributing to the excess in the $K^+K^-$ invariant mass 
spectrum seen around 1.5 GeV/$c^2$.

  We find that effects of interference between different two-body 
intermediate states can have significant influence on the observed 
two-particle mass spectra and a full amplitude analysis of three-body 
$B$ meson decays is required for a more complete understanding.
This will be possible with increased statistics.

\section*{Acknowledgement}

We wish to thank the KEKB accelerator group for the excellent
operation of the KEKB accelerator. We acknowledge support from the
Ministry of Education, Culture, Sports, Science, and Technology of 
Japan and the Japan Society for the Promotion of Science; 
the Australian Research Council and the Australian Department of 
Industry, Science and Resources; the Department of Science and 
Technology of India; the BK21 program of the Ministry of Education 
of Korea and the CHEP SRC program of the Korea Science and Engineering
Foundation; the Polish State Committee for Scientific Research under 
contract No.2P03B 17017; the Ministry of Science and Technology of 
Russian Federation; the National Science Council and the Ministry of 
Education of Taiwan; and the U.S. Department of Energy.

%%%%%%%%%%%%%%%%%%%%%%%%%%%%%%%%%%%%%%%%%%%%%%%%%%%%%%%%%%%%%%%%%%%%%%
%%%%%%%%%%%%%%%%%%%%%%%%   References   %%%%%%%%%%%%%%%%%%%%%%%%%%%%%%
%%%%%%%%%%%%%%%%%%%%%%%%%%%%%%%%%%%%%%%%%%%%%%%%%%%%%%%%%%%%%%%%%%%%%%

\clearpage
%%%%%%%%%%%%%%%%%%%%%%%%%%%%%%%%%%%%%%%%%%%%%%%%%%%%%%%%%%%%%%%%%%%%%%%%
%%%%%%%%%%%%%%%%%%%%%%%%%%%%%%   TABLES   %%%%%%%%%%%%%%%%%%%%%%%%%%%%%%
%%%%%%%%%%%%%%%%%%%%%%%%%%%%%%%%%%%%%%%%%%%%%%%%%%%%%%%%%%%%%%%%%%%%%%%%

\begin{table}[t]              % Table I
  \caption{Branching fractions and 90\% C.L. upper limits for
           $B^+\to K^+h^+h^-$ final states.}
  \medskip
  \label{results1}
  \begin{tabular}{lccc} \hline \hline
  Three-body       &   Efficiency   &   Yield    &  ${\cal{B}}$  \\
  mode             &      (\%)      &  (events)  &  $(10^{-6})$  \\ \hline
  $K^+\pi^+\pi^-$  & $17.3$ &  $237\pm23$   & $55.6\pm5.8\pm7.7$ \\
  $K^+K^+K^-$      & $24.0$ &  $210\pm21$   & $35.3\pm3.7\pm4.5$ \\
  $K^-\pi^+\pi^+$  & $16.2$ &   $12\pm9$    & $<7.0$             \\
  $K^+K^+\pi^-$    & $14.2$ &  $2.0\pm5.3$  & $<3.2$             \\
  $K^+K^-\pi^+$    & $14.6$ &   $26\pm12$   & $<12$              \\
\hline \hline
  \end{tabular}
\end{table}
\begin{table*}[t]              % Table II
\caption{Results of the fit to the $\Delta E$ distribution for different
         regions in the $K^+\pi^+\pi^-$ Dalitz plot. Columns list the 
         definition of each region, reconstruction efficiency from Monte 
         Carlo simulation and signal yield.}
\medskip
\label{kpp_fits}
  \begin{tabular}{lccr}  \hline \hline
   Dalitz plot  & Mass range   &  Efficiency  & Yield     \\
   region       & (GeV/$c^2$)  &     (\%)     & (events)  \\ \hline
 I & $M(K\pi)<1.00$; $M(\pi\pi)>1.50$ & $20.7\pm3.7$ 
                   & $47.1^{+9.3}_{-8.7}$                 \\
 II  & $1.00<M(K\pi)<2.00$; $M(\pi\pi)>1.50$ 
                   & $19.2\pm1.3$ & $56.0^{+12}_{-11}$    \\
 III & $M(\pi\pi)<0.90$; $M(K\pi)>2.00$
                   & $16.7\pm2.4$ & $17.7^{+8.4}_{-7.8}$  \\
  IV & $0.90<M(\pi\pi)<1.06$; $M(K\pi)>2.00$
                   & $19.9\pm3.2$ & $34.7^{+7.9}_{-7.3}$  \\
  V  & $1.06<M(\pi\pi)<1.50$; $M(K\pi)>2.00$ 
                   & $19.6\pm1.7$ & $33.4^{+8.8}_{-8.2}$  \\
 VI  & $M(K\pi)<2.00$; $M(\pi\pi) < 1.50$ 
                   & $14.7\pm3.3$ & $14.9^{+6.3}_{-5.7}$  \\
 VII & $M(K\pi)>2.00$; $M(\pi\pi) > 1.50$
                   & $16.1\pm0.5$ & $11.8^{+9.4}_{-8.8}$  \\ 
\hline \hline 
  \end{tabular}
\end{table*}
\begin{table*}[htb]              % Table III
\caption{Results of the fit to the $\Delta E$ distribution for different
         regions in the $K^+K^+K^-$ Dalitz plot. Columns list the 
         definition of each region, reconstruction efficiency from Monte
         Carlo simulation and signal yield.}
\medskip
\label{kkk_fits}
  \begin{tabular}{lccr}  \hline \hline
 Dalitz plot &  Mass range      & Efficiency
             &  Yield~~                                  \\
  ~~~region  &  (GeV/$c^2$)     & (\%)  
             &  (events)                                 \\ \hline
 I   & $M(KK)_{\rm min}<1.04$ 
               & $24.6\pm2.5$ & $35.7^{+7.1}_{-6.5}$     \\
 II  & $1.04<M(KK)_{\rm min}<2.00$
               & $23.3\pm0.8$ & $113^{+15}_{-14}$        \\
 III & $M(KK)_{\rm min}>2.00$; $M(KK)_{\rm max}>3.40$
               & $23.9\pm1.1$ & $14.7^{+7.0}_{-6.4}$     \\
 IV  & $M(KK)_{\rm min}>2.00$; $M(KK)_{\rm max}<3.40$
               & $24.7\pm0.8$ & $32.3^{+6.8}_{-6.1}$     \\
\hline \hline
  \end{tabular}
\end{table*}
\begin{table*}[htb]              % Table IV
\caption{List of systematic errors (in percent) for the
         $B^+\to K^+h^+h^-$ branching fractions.}
\medskip
\label{khh_syst}
  \begin{tabular}{lccccc}  \hline \hline
  Source &$K^+\pi^+\pi^-$  & $K^-\pi^+\pi^+$  & $K^+K^-\pi^+$  &  $K^+K^+\pi^-$  &  $K^+K^+K^-$  \\ \hline 

 $B\to D\pi$ and $D\to K\pi$ branching fractions & 
                                 9.7 & 9.7 & 9.7 & 9.7 & 9.7   \\
 efficiency nonuniformity over the Dalitz plot   & 
                                 7.6 &  -  &  -  &  -  & 3.7   \\
 background and signal parameterization          & 
                                 6.3 &  -  &  -  &  -  & 4.7   \\
 particle identification                         & 
                                  -  &  -  & 3.0 & 3.0 & 6.0   \\ \hline
    total         &           13.8 &  9.7 & 10.1 & 10.1 & 12.9 \\
\hline \hline
  \end{tabular}
\end{table*}
\begin{table*}[htb]              % Table V
\caption{Results of the simultaneous fit to the $K^+\pi^+\pi^-$ 
         final state.}
\medskip
\label{kpp_min}
  \begin{tabular}{lcccc}  \hline \hline
    Two-body        & Efficiency   & Yield
                    & Significance
    & ${\cal{B}}_{B^+\to Rh^+}\times{\cal{B}}_{R\to h^+h^-}$  \\

   ~~~mode          & (\%)         & (events) 
                    & $(\sigma)$ 
    & $(10^{-6})$  \\ \hline

 $K^*(892)^0\pi^+$ & 18.9 & $60^{+13}_{-12}$  & 6.2
                    & $12.9^{+2.8+1.4+2.3}_{-2.6-1.4-4.5}$    \\
  $K_X(1400)\pi^+$  & 16.2 & $58^{+14}_{-13}$  & 4.9
                    & $14.5^{+3.5+1.8+3.3}_{-3.3-1.8-6.5}$    \\
 $\rho^0(770)K^+$   & 15.1 & $9^{+13}_{-12}$ & 0.8 & $<12$    \\
   $f_0(980)K^+$    & 17.8 & $42^{+11}_{-10}$  & 5.0
                    & $9.6^{+2.5+1.5+3.4}_{-2.3-1.5-0.8}$     \\
  $f_X(1300)K^+$    & 16.9 & $46^{+14}_{-13}$  & 3.9
                    & $11.1^{+3.4+1.4+7.2}_{-3.1-1.4-2.9}$    \\
\hline \hline
  \end{tabular}
\end{table*}
\begin{table*}[htb]              % Table VI
\caption{Results of the simultaneous fit to the $K^+K^+K^-$ final state.}
\medskip
\label{kkk_min}
  \begin{tabular}{lcccc}  \hline \hline
    Two-body        & Efficiency   & Yield
                    & Significance
    & ${\cal{B}}_{B^+\to Rh^+}\times{\cal{B}}_{R\to h^+h^-}$  \\

   ~~~mode          & (\%)         & (events) 
                    & $(\sigma)$ 
    & $(10^{-6})$  \\ \hline 

  $\phi(1020)K^+$   & 23.6 & $42^{+8.7}_{-7.9}$ & 7.2
                    & $7.2^{+1.5+0.9+0.4}_{-1.4-0.9-0.4}$ \\
  $f_X(1500)K^+$    & 21.3 & $146^{+17}_{-17}$  & 12 
                    & $27.6^{+3.2+3.5+1.4}_{-3.2-3.5-1.4}$ \\
\hline \hline
  \end{tabular}
\end{table*}

\clearpage
%%%%%%%%%%%%%%%%%%%%%%%%%%%%%%%%%%%%%%%%%%%%%%%%%%%%%%%%%%%%%%%%%%%%%%%%
%%%%%%%%%%%%%%%%%%%%%%%%%%%%%   FIGURES   %%%%%%%%%%%%%%%%%%%%%%%%%%%%%%
%%%%%%%%%%%%%%%%%%%%%%%%%%%%%%%%%%%%%%%%%%%%%%%%%%%%%%%%%%%%%%%%%%%%%%%%


%%%%%%%%%%%%%%%%%%%%%%%%%%%%%%%%%%%%%%%%%%%%%%%%%%%%%%%%%%%%%%%%%%%%%%%%
%%%%%%%%%%%%%%%%%%%%%%%%%%%%%%%%%%%%%%%%%%%%%%%%%%%%%%%%%%%%%%%%%%%%%%%%
%%%%%%%%%%%%%%%%%%%%%%%%%%%%%%%%%%%%%%%%%%%%%%%%%%%%%%%%%%%%%%%%%%%%%%%%


\begin{thebibliography}{99}
\bibitem{dEreg}{The negative $\Delta E$ region could contain events 
        from $B\to Khh\pi$ with an additional low momentum neutral or 
        charged pion that is not included in the $M_{\rm bc}$ and 
        $\Delta E$ computation. Therefore, we do not use the 
        $\Delta E<-0.1$~GeV region for sideband studies and also 
        exclude it from the fits.}
\end{thebibliography}
\end{document}